\documentclass[twocolumn]{aastex631}

\newcommand{\Msun}{${\rm M}_{\odot}$}
\newcommand{\Rsun}{${\rm R}_{\odot}$}

\newcommand{\kms}{km\,s$^{-1}$}

\newcommand{\NaI}{Na~{\sc i}}

\newcommand{\OI}{O~{\sc i}}

\newcommand{\OIV}{O~{\sc iv}}
\newcommand{\OV}{O~{\sc v}}

\newcommand{\HI}{H~{\sc i}}

\newcommand{\CIII}{C~{\sc iii}}
\newcommand{\CIV}{C~{\sc iv}}

\newcommand{\NIV}{N~{\sc iv}}
\newcommand{\NIII}{N~{\sc iii}}
\newcommand{\NV}{N~{\sc v}}

\newcommand{\HeI}{He~{\sc i}}
\newcommand{\HeII}{He~{\sc ii}}

\def\apjl{Astrophys.~J.}
\def\apjs{Astrophys.~J.~Suppl.}

\def\aap{Astron.~Astrophys.}

\received{June 1, 2024}
\revised{June 27, 2024}
\accepted{June 27, 2024 by ApJL}

\shorttitle{SN 2024ggi}
\shortauthors{Zhang et al.}
\graphicspath{{./}{figures/}}

\begin{document}
\title{Probing the Shock Breakout Signal of SN 2024ggi from the Transformation of Early Flash Spectroscopy}

\author[0000-0002-8296-2590]{Jujia Zhang}\thanks{E-mail:jujia@ynao.ac.cn}
\affiliation{Yunnan Observatories (YNAO), Chinese Academy of Sciences (CAS), Kunming, 650216, China}
\affiliation{International Centre of Supernovae, Yunnan Key Laboratory, Kunming 650216, China}
\affiliation{Key Laboratory for the Structure and Evolution of Celestial Objects, CAS, Kunming, 650216, China}

\author[0000-0003-0599-8407]{Luc Dessart}
\affiliation{Institut d'Astrophysique de Paris, CNRS-Sorbonne Universit{\'e}, 98 bis boulevard Arago, F-75014 Paris, France}

\author[0000-0002-7334-2357]{Xiaofeng Wang}
\affiliation{Physics Department, Tsinghua University, Beijing, 100084, China}

\author{Qian Zhai}
\affiliation{Yunnan Observatories (YNAO), Chinese Academy of Sciences (CAS), Kunming, 650216, China}
\affiliation{Key Laboratory for the Structure and Evolution of Celestial Objects, CAS, Kunming, 650216, China}

\author[0000-0002-6535-8500]{Yi Yang} 
\affiliation{Physics Department, Tsinghua University, Beijing, 100084, China}
\affiliation{Department of Astronomy, University of California, Berkeley, CA 94720-3411, USA}

\author{Liping Li}
\affiliation{Yunnan Observatories (YNAO), Chinese Academy of Sciences (CAS), Kunming, 650216, China}
\affiliation{International Centre of Supernovae, Yunnan Key Laboratory, Kunming 650216, China}
\affiliation{Key Laboratory for the Structure and Evolution of Celestial Objects, CAS, Kunming, 650216, China}

\author{Han Lin}
\affiliation{Yunnan Observatories (YNAO), Chinese Academy of Sciences (CAS), Kunming, 650216, China}
\affiliation{International Centre of Supernovae, Yunnan Key Laboratory, Kunming 650216, China}
\affiliation{Key Laboratory for the Structure and Evolution of Celestial Objects, CAS, Kunming, 650216, China}


\author{Giorgio Valerin}
\affiliation{INAF-Osservatorio Astronomico di Padova, Vicolo dell'Osservatorio 5, 35122 Padova, Italy}

\author{Yongzhi Cai}
\affiliation{Yunnan Observatories (YNAO), Chinese Academy of Sciences (CAS), Kunming, 650216, China}
\affiliation{International Centre of Supernovae, Yunnan Key Laboratory, Kunming 650216, China}
\affiliation{Key Laboratory for the Structure and Evolution of Celestial Objects, CAS, Kunming, 650216, China}

\author[0000-0003-0292-4832]{Zhen Guo}
\affiliation{Instituto de F{\'i}sica y Astronom{\'i}a, Universidad de Valpara{\'i}so, ave. Gran Breta{\~n}a, 1111, Casilla 5030, Valpara{\'i}so, Chile}
\affiliation{Centre for Astrophysics Research, University of Hertfordshire, Hatfield AL10 9AB, UK}
\affiliation{Millennium Institute of Astrophysics,  Nuncio Monse{\~n}or Sotero Sanz 100, Of. 104, Providencia, Santiago,  Chile}

\author[0000-0002-1094-3817]{Lingzhi Wang}
\affiliation{Chinese Academy of Sciences South America Center for Astronomy (CASSACA), National Astronomical Observatories, CAS, Beijing, China}

\author{Zeyi Zhao}
\affiliation{Yunnan Observatories (YNAO), Chinese Academy of Sciences (CAS), Kunming, 650216, China}
\affiliation{International Centre of Supernovae, Yunnan Key Laboratory, Kunming 650216, China}
\affiliation{Key Laboratory for the Structure and Evolution of Celestial Objects, CAS, Kunming, 650216, China}

\author{Zhenyu Wang}
\affiliation{Yunnan Observatories (YNAO), Chinese Academy of Sciences (CAS), Kunming, 650216, China}
\affiliation{International Centre of Supernovae, Yunnan Key Laboratory, Kunming 650216, China}
\affiliation{Key Laboratory for the Structure and Evolution of Celestial Objects, CAS, Kunming, 650216, China}

\author{Shengyu Yan} 
\affiliation{Physics Department, Tsinghua University, Beijing, 100084, China}



\begin{abstract}
We present early-time, hour-to-day cadence spectroscopy of the nearby type II supernova (SN II) 2024ggi, which was discovered at a phase when the SN shock just emerged from the red-supergiant (RSG) progenitor star. Over the first few days after the first light, SN\,2024ggi exhibited prominent narrow emission lines formed through intense and persistent photoionization of the nearby circumstellar material (CSM). In the first 63 hours, spectral lines of He, C, N, and O revealed a rapid rise in ionization, 
as a result of the progressive sweeping-up of the CSM by the shock. 
The duration of the IIn-like spectra indicates a dense and relatively confined CSM distribution extending up to $\sim\,4\,\times\,10^{14}$\,cm. Spectral modeling reveals a CSM mass loss rate at this region exceeding $5 \times\, 10^{-3}$\,\Msun\,yr$^{-1}$\,is required to reproduce low-ionization emissions, which dramatically exceeds that of an RSG. Analyzing H$\alpha$ emission shift implies the velocity of the unshocked outer CSM to be between 20 and 40 \kms, matching the typical wind velocity of an RSG. The differences between the inner and outer layers of the CSM and an RSG progenitor highlight a complex mass loss history before the explosion of SN 2024ggi.

\end{abstract}



\section{Introduction} 
\label{sec:intro}
Type II supernovae (SNe), characterized by the presence of prominent hydrogen features in their spectra \citep{1997ARA&A..35..309F,2017hsn..book..195G} are the death throes of most massive stars (e.g. red supergiant, RSG, \citealp{2003ApJ...591..288H}). A fraction of SNe II exhibits interaction signatures with circumstellar material (CSM), characterized by narrow optical emission lines with broad electron-scattering wings classified as Type IIn SNe \citep{1985ApJ...289...52N,1990MNRAS.244..269S}.
Some of the diversity of SNe IIn lies in the duration of the interaction signatures, which can range from a few days (e.g., SN 2013fs, \citealp{2017NatPh..13..510Y}) to several weeks (e.g., SN 1998S, \citealp{2000ApJ...536..239L,2001MNRAS.326.1448C,2001MNRAS.325..907F,2015ApJ...806..213S,2016MNRAS.458.2094D}), months (e.g., SN 2010jl, \citealp{2012AJ....144..131Z,2014ApJ...797..118F}), and even a few years (e.g., SN 2015da, \citealp{2020A&A...635A..39T}). This indicates a range in spatial scales and mass for the CSM around their progenitor stars.

These cataclysmic events provide a unique window into the final stages of stellar evolution, especially the mass-loss history and the environment surrounding the progenitor stars. The mass-loss rates deduced from SN observations often differ from those derived from studies of RSG in their quiescent phases, underscoring the gaps in our understanding of late-stage stellar evolution. 

For example, the variety of SNe II with short-lived IIn-like spectral features, as witnessed in supernovae like SN 2013fs \citep{2017A&A...603A..51D,2017NatPh..13..510Y}, SN 2014G \citep{2016MNRAS.462..137T}, SN 2017ahn \citep{2021ApJ...907...52T}, SN 2018zd \citep{2020MNRAS.498...84Z,2021NatAs...5..903H}, SN 2020pni \citep{2022ApJ...926...20T}, 2020tlf \citep{2022ApJ...924...15J}, and the recent SN 2023ixf (e.g., \citealp{2023ApJ...956L...5B,Hiramatsu2023ApJ...955L...8H,2023ApJ...956...46S,2023SciBu..68.2548Z,2024Natur.627..759Z}), reveals the diversity of CSM structures and the mass-loss histories of that subset of SNe and their associated massive star progenitors. Quantitative studies show that over 30\% of SNe II exhibit these transient IIn-like features (e.g., \citealp{2021ApJ...912...46B,2023ApJ...952..119B,2024arXiv240302382J}), suggesting that such mass loss events are common in massive stars about to explode, with certain SNe displaying intermediate properties between regular SNe II and strongly interacting SNe IIn.

Moreover, the early IIn-like spectra resulting from ionizations of CSM by shock photons 
provide a way to investigate the initial shock breakout (SBO) signal of SN explosion. The earliest emission of electromagnetic radiation from an SN explosion is associated with SBO, a brief yet brilliant event signifying the transition from an opaque to a transparent state as the expanding shockwave reaches the stellar surface \citep{2017hsn..book..967W}. Once the shock approaches the progenitor (within an optical depth of $\tau \sim 10-30$), radiation begins to leak from the shock, initiating the ionization of the cool atmosphere and environment of the RSG.

The ionization process, occurring on timescales from seconds to tens of minutes, is marked by a UV flash \citep{2017hsn..book..967W}. It is followed by UV and optical emissions from the cooling envelope. However, if there is an optically thick CSM or dust shell, the SBO might occur within this shell, and the shock emission will be reddened and prolonged, as seen in SN 2023ixf \citep{2023ApJ...956L...5B,2024Natur.627..754L,2024Natur.627..759Z}.

In this letter, we present the spectroscopic observations of a nearby SN II in NGC 3621, which provides another chance to detect the SBO signals. SN 2024ggi was discovered by ATLAS (Asteroid Terrestrial-impact Last Alert System, \citealp{2024TNSTR1020....1T,2024arXiv240609270C}) on April 11.14, 2024 (UTC dates are used throughout this paper) in the nearby galaxy NGC 3621, located at a distance of D = 7.0 $\pm$ 0.2 Mpc (based on the averaged Cepheid distance derived by different period-luminosity relations presented in \citealp{kanbur2003extra}). The archive images taken with the Hubble Space Telescope about 20 years before the explosion reveal an RSG progenitor with a temperature $T_{\star}=3290_{-27}^{+19}$~K and radius $R_{\star}=887_{-51}^{+60}$~\Rsun\,\citep{2024ApJ...969L..15X}. With the first light date at MJD = 60411.03 $\pm$ 0.05 determined through high-cadence photometric observations (Yan et al. in prep.), SN\,20024ggi stands out as an SN II with extremely early photometric and spectroscopic observations \citep{2024TNSTR1020....1T, 2024TNSAN.104....1Z}.

\begin{figure*}
\centering
\includegraphics[width=18cm,angle=0]{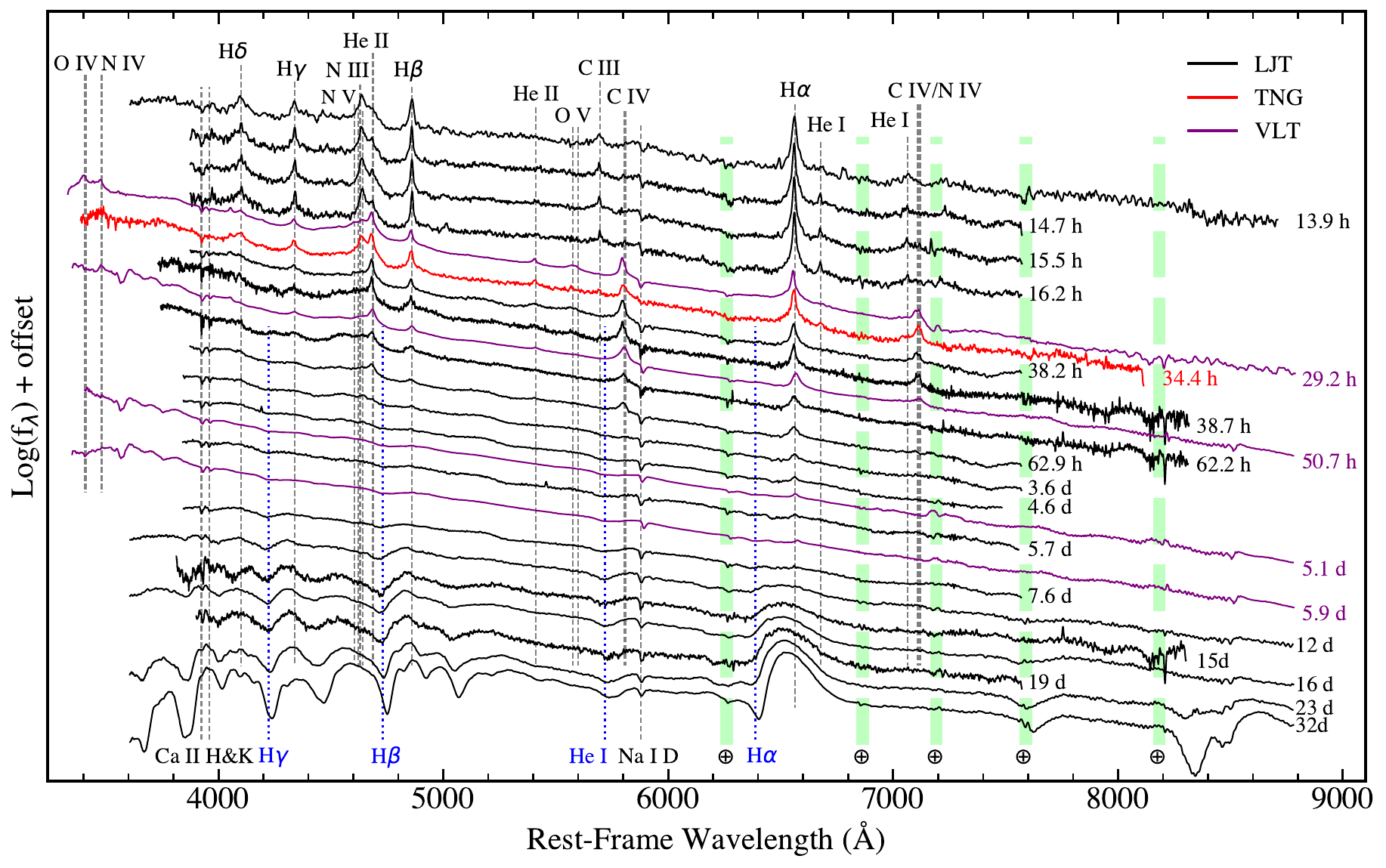}
 \caption{Spectral evolution of SN 2024ggi. Dashed grey lines indicate the rest-frame wavelengths of spectral features, while dotted blue lines represent a blueshift of $-$8000 \kms. Artifacts resulting from the incomplete removal of telluric absorption and skyline emissions are identified by dash-dotted green lines, with each marked with an Earth symbol indicating the terrestrial origin. Details of the observations, including dates and conditions, are given in Table \ref{Tab:Spec_log}. The spectra are provided in machine-readable format as Data behind the Figure and can be downloaded in the online version of this article.}

\label{<spe_whole>}
\end{figure*}

\section{Spectral observation}
\label{sec:specobs}

\subsection{Classification}
\label{subsec:clas}
Utilizing the Li-Jiang 2.4-m telescope (LJT; \citealp{2015RAA....15..918F}) equipped with YFOSC (Yunnan Faint Object Spectrograph and Camera; \citealp{2019RAA....19..149W}), we obtained a classification spectrum for SN 2024ggi at $\sim$ 13.9 hours after its explosion \citep{2024TNSAN.104....1Z}. This classification spectrum was predominantly characterized by narrow emission lines of H, He, and CNO elements, as seen in Figure \ref{<spe_whole>}. These features, also known as flash features \citep{2014Natur.509..471G}, are commonly observed in SNe IIn (e.g., SN 1998S, \citealp{2000ApJ...536..239L}). They are generated by recombinations of the surrounding dense CSM that is photoionized by radiation from the embedded shock.

\begin{figure*}
\centering
\includegraphics[width=16cm,angle=0]{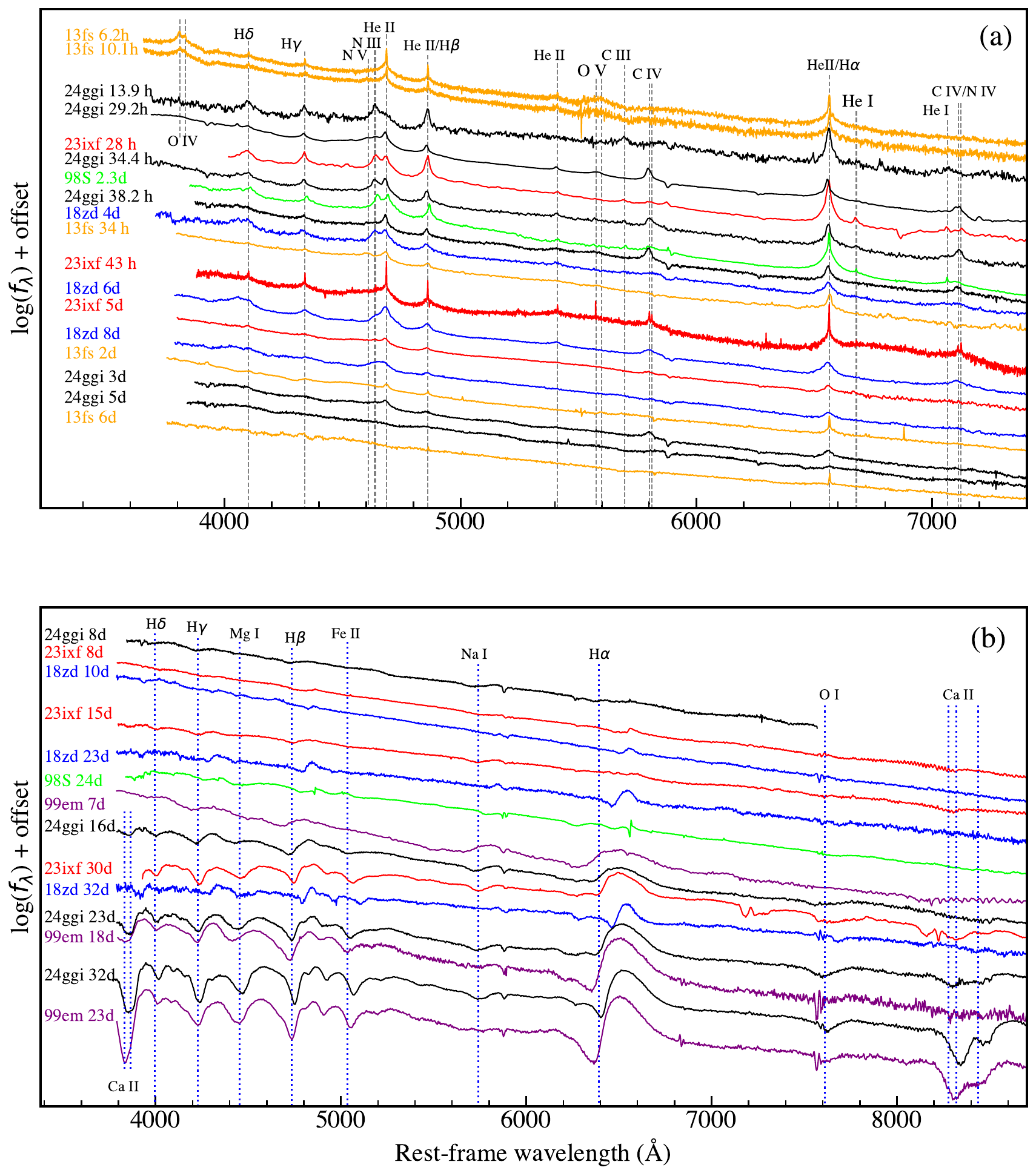}
\caption{Spectral comparison between SN 2024ggi and the representative SNe II. (a) Flash spectroscopy of SN 2024ggi, compared with those of SN 1998S \citep{2000ApJ...536..239L,2001MNRAS.325..907F,2015ApJ...806..213S}, SN 2013fs \citep{2017NatPh..13..510Y}, SN 2018zd \citep{2020MNRAS.498...84Z}, and SN 2023ixf \citep{2023SciBu..68.2548Z}.  (b) Post-flash spectra of SN 2024ggi, 
compared with those from SN 1998S, SN 1999em \citep{2001ApJ...558..615H,2002PASP..114...35L,2006A&A...447..691D}, SN 2018zd, and SN 2023ixf. The extinction correction was applied to each spectrum, for which the extinction of SN 2024ggi is derived in Section \ref{subsec:ext} while the extinction values of the 
comparisons are adopted from the lectures. Dashed grey lines represent the rest-frame wavelengths of spectral features while the dotted blue lines represent a blueshift of $-$8000 km s$^{-1}$ in velocity space. }
\label{<spe_comp>}
\end{figure*}

Interestingly, in the initial-phase spectrum, SN 2024ggi exhibits a significant difference from other SNe II-P/CSM events like SN 2013fs, SN 2018zd, and SN 2023ixf. As presented in Figure \ref{<spe_comp>} (a), the classification spectrum of SN 2024ggi lacks strong emission features of high ionization states. It showed lines of \HeI, \NIII, and \CIII\,rather than those of \HeII, \NV\,and \CIV\,or highly-ionized \OV\,$\lambda$$\lambda$\,5576,5598 and \NV\,$\lambda$$\lambda$\,4604,4620 lines that appeared in SN 2013fs. 

In particular, SN 2024ggi and SN 2013fs exhibit two opposite spectral line morphologies in the 4600-4700 \AA\,region. The spectra of SN 2013fs at $t\leq 10$ hr (where $t$ denotes time after the first light) show a narrow \HeII\,$\lambda$\,4686 emission line, while the nearby \NIII\,$\lambda$$\lambda$\,4630,4641 doublets appear flatter without prominent narrow-line components. Instead, there is a small bump consisting of \NV\,$\lambda$$\lambda$\,4604,4620. However, the classification spectrum of SN 2024ggi reveals narrow emission lines of \NIII\,$\lambda$$\lambda$\,4630,4641, while the position corresponding to \HeII\,$\lambda$\,4686 appears flat. These two SNe belong to those with the earliest discoveries among the sample of SNe II-P with CSM. Spectroscopic data from other samples, usually obtained one day after the SBO, did not show similar phenomena, suggesting a high dependency of early-time spectra on the exact post-breakout epoch and the CSM 
properties.

\subsection{High-cadence sampling within the initial 72 hours}
\label{subsec:72h}
Given the observed IIn-like features in the classification spectrum and a relatively lower ionization state revealed by these features, we initiated an hour-cadence observation campaign with the LJT. A total of four spectra were acquired within 2.4 hours after the identification, which allowed us to monitor possible variations in the ionization state for this young SN II.

\begin{figure}
\centering
\includegraphics[width=8.5cm,angle=0]{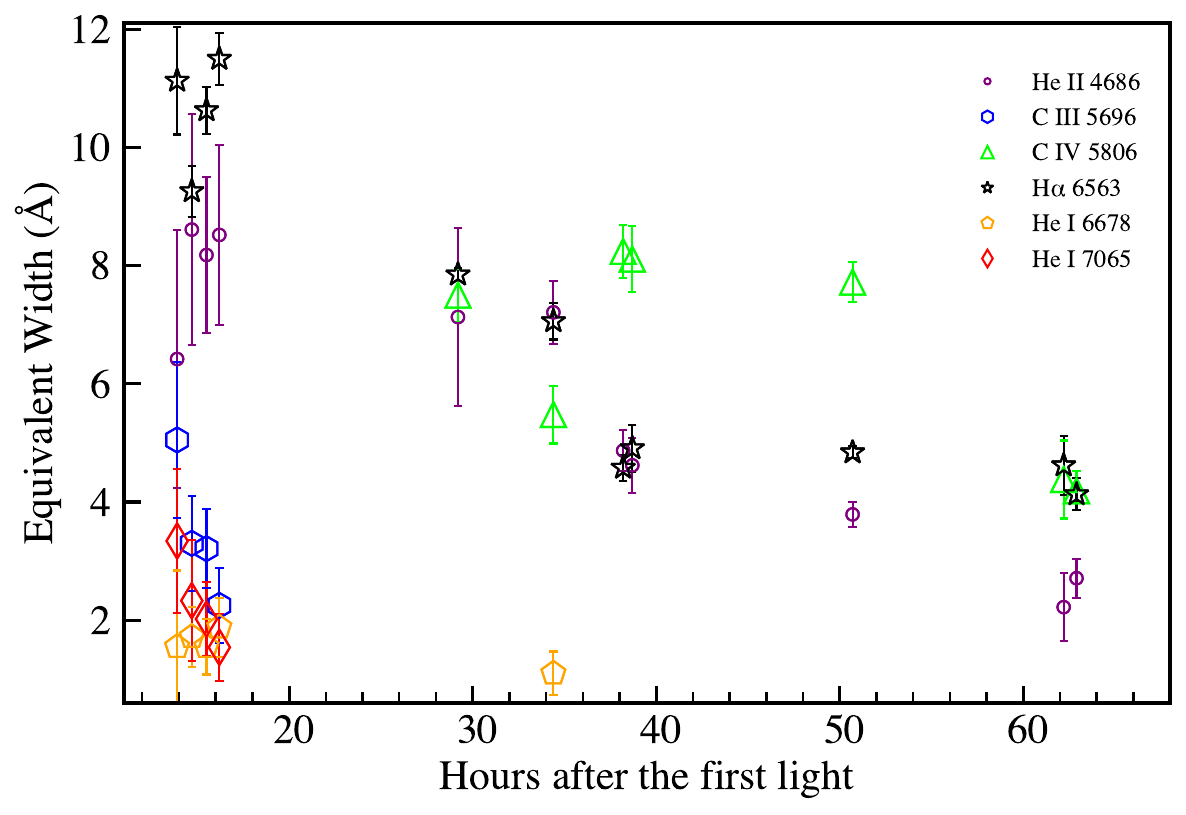}
\caption{Equivalent width evolution of ionization line in SN 2024ggi at $t<3$ d. \CIV\,5806 is the abbreviation of \CIV\,$\lambda$$\lambda$\, 5801, 5811. }
\label{<FI>}
\end{figure}

From the second spectrum, SN 2024ggi exhibited narrow \HeII\,$\lambda$\,4686 emission that was absent before. It is possible that, in the first spectrum, the CSM has not been shocked enough to generate sufficient heat for emitting a significant amount of \HeII, except for the deep CSM near the shock, where the material could be rushing. The subsequent two spectra did not show significant morphological evolution. Further analysis of the equivalent widths (EWs) of the main spectral lines, as shown in Figure \ref{<FI>}, reveals that within the 1.5 hr from $t\sim$14.7 hr to $t\sim$ 16.2 hr, the strength of \HeII\,$\lambda$\,4686 and \HeI\,$\lambda$\,6678 lines remained almost unchanged. In contrast, the \CIII\,$\lambda$\,5696 and \HeI\,$\lambda$\,7065 lines beaome gradually weak. 

On the second day after explosion, an observation relay across different time zones was conducted with LJT, Telescopio Nazionale Galileo (TNG), and Very Large Telescope Unit 1 (VLT UT-1), which allowed a continuous monitoring of the rapid evolution of SN 2024ggi. In particular, to examine the structure of narrow spectral lines, we utilized the cross-dispersion capability of YFOSC with a resolution of 3500.
This resolution had proven its effectiveness in our previous research of SN 2023ixf, enabling us to discern the intricate structure of H$\alpha$ and provide a more precise constraint on its broadening \citep{2023SciBu..68.2548Z}. 

Nonetheless, in the case of SN 2024ggi, the H$\alpha$ line had already undergone significant broadening in the spectra taken at $t\sim 38.7$ hr, with a full width at half maximum (FHWM) of $\sim$ 700 \kms, exceeding the instrumental FWHM (i.e., $\sim$ 85 \kms). Concurrently, as the narrow line broadened swiftly, its equivalent width (EW) decline steeply at $t \gtrsim 30$ hr, as illustrated in Figure \ref{<FI>}.

The rapid broadening and decreasing of the H$\alpha$ narrow component testifies the rapid spectral evolution of SN 2024ggi. By $t\sim$ 29 hr, many highly ionized spectral lines emerged in the spectrum, including \OIV, \NIV, and \CIV, and even \OV. Due to limitations in the signal-to-noise ratio (S/N) of the spectrum, the \HeII\,$\lambda$\,5411 line can only be marginally detected at $t < 1$ d, but it becomes visible at $t\sim$ 29 hr. 
The most notable change is that 
One day after explosion, the emission lines of \HeI\,$\lambda$\,7065 and \CIII\,$\lambda$\,5696 disappeared in the spectra, but replaced by \CIV\,$\lambda$$\lambda$\,5801, 5811, \CIV\,$\lambda$\,7110 and \NIV\,$\lambda$\,7122 lines. As shown in Figure \ref{<FI>}, \HeII\,$\lambda$\,4686 continues to weaken throughout the day, while the \CIV\,$\lambda$$\lambda$\,5801, 5811 progressively gain the strengthens. We notice that the narrow \HeI\,$\lambda$\,6678 line disappeared at 29.2 hr but it reappeared five hours later. The \HeI\,$\lambda$\,5876 line is not detected perhaps due to that it coincides with the  \NaI\,D absorption of the Milky Way.

At $t\sim 29$ hr, the spectral line flux and profile morphology in this region are unreliable due to a saturation in the wavelength range from $\sim$4610\AA\,to $\sim$4700 \AA. Therefore, we cannot confirm whether the appearance of \NV\,at this time is genuine, nor can we ascertain the reliability of the intensity contrast between \NIII\,and \HeII. At $t\sim$ 34 hr, the narrow lines of \HeII\,$\lambda$\,4686 and \NIII\,$\lambda$$\lambda$\,4630,4641 have comparable intensities and thus formed a double peak profile. Such a double-peak structure has been also seen in some SNe II. For instance, it appeared in SN 2023ixf at $t\sim$ 1.2 d, SN 1998S at $t\sim$ 2.3 d, and SN 2018zd at $t\sim$ 4 d. SN 2013fs may also exhibit a similar structure at $t\sim$ 1.4 d, except that the left peak corresponds to \NV\,$\lambda$$\lambda$\,4604,4620 instead of \NIII\,in its spectrum \citep{2017A&A...603A..51D,2017NatPh..13..510Y}. The \NV\,doublet is also visible in the spectra of SN 2024ggi taken at $t=$ 29.2h, 38.3h, and 38.7h. 
This indicates that SN 2013fs and SN 2024ggi can reach similar ionization levels, though they may have different shock emissions and CSM environments.

At $t>$ 38 hr, \NIII\,doublet becomes weaker than \HeII\,$\lambda$\,4686 in SN 2024ggi. However, the \NIII\,doublet remains unchanged over the following days while \HeII\, shows a faster decline in intensity. Overall, all narrow emission lines become less distinct after the third day, as they broaden significantly with a velocity exceeding 1000 \kms. The disappearance of narrow emission lines indicates that the IIn-like phase is ending and SN 2024ggi is gradually entering a new stage of evolution.

The rapid spectral evolution relates in part to a temperature change. We can roughly estimate the temperature of the SN photosphere through fits to the spectral energy distribution (SED) assuming a blackbody emission. In Figure \ref{<bolo>}, the temperature of SN 2024ggi is inferred to be about 13500 K during the phase from 13.9 hr to 16.2 h, and it increases sharply from $\sim$\,13500 K to $\sim$\,26000 K about half a day later (also reported by \citealp{2024arXiv240507964C}). The rapid temperature rise is concomitant with the rapid increase in ionization, and both are attributed to the photoionizing radiation from the shock.

\subsection{Flash-to-Photospheric Phase Transition}
\label{Photosp}
Since SN 2024ggi exhibited a rapid evolution, we maintained frequent monitoring until the third day, gradually tapering off the pace after that. As presented in Figure \ref{<spe_whole>}, we can only see a few weak `narrow' emission lines, such as H$\alpha$ and \NIII\,$\lambda$$\lambda$\,4634,4641 doublet at $\sim$ 3.6 d. Meanwhile, P-cygni profiles of H$\beta$, H$\gamma$, and \HeI\,$\lambda$\,5876 begin to emerge. 

At $3 < t < 15$ d, the spectral lines of SN 2024ggi undergo a significant transition. Initially, these lines exhibited broadening due to electron scattering, formed within the unshocked CSM. However, they gradually shift to Doppler broadening as they initially formed within the fast-moving dense shell and subsequently arises also from the ejecta as the dense shell gradually becomes optically thinner. This evolution is accompanied by a noticeable blueshift in their emission peaks. Additionally, these lines begin to appear as P-Cygni profiles, with both absorption and emission components increasing in strength over time.

Roughly two weeks after the explosion, SN 2024ggi developed spectral features typical of SNe IIP in both optical and NIR spectra, i.e., the appearance of broad \HI\,and \HeI\,lines. For example, as seen in Figure\ref{<NIR>}, the NIR spectrum of SN 2024ggi, obtained with the 6.5~m Magellan telescope equipped with FIRE (Folded-port InfraRed Echellette) at t$\sim$ 14 d, reveals similar spectral features as seen in SN2017eaw \citep{2019ApJ...876...19S}.

As observed in SN 2018zd \citep{2020MNRAS.498...84Z} and SN 2023ixf \citep{2023SciBu..68.2548Z}, the SN-CSM interaction creates a slow evolution in spectral features after the narrow, electron-scattering broadened emission lines disappear (Figure \ref{<spe_comp>}). During shock wave propagation, the CSM is shocked, and compressed into a dense shell, converting a fraction of kinetic energy into thermal energy and heating the CSM as well as boosting the SN luminosity. In the case of SN 2024ggi, photoionization predominantly facilitates this heating process within the first three days. Consequently, emission from the post-shock gas, such as the cold dense shell (CDS), originates from the release of shock-deposited energy (i.e., deposited at earlier times). The relatively faint spectral lines observed approximately five days later are likely attributed to the spectrum primarily forming within the CDS during this period. This weak emission is indicative of a steep density gradient of the CDS \citep{2017A&A...603A..51D}.

We note that the evolution speed of the photospheric features in SN 2024ggi is slower than that of the typical SN 1999em but still faster than in SN 2018zd and SN 2023ixf. As illustrated in Figure \ref{<spe_comp>} (b), SN 2024ggi evolves into spectra with prominent P-Cygni profiles at around $t\sim$ 23 d, and its spectrum at $t\sim 32$ d shows a close resemblance to that of a regular SN IIP. Nevertheless, SN 2024ggi evolves faster in the first month, regardless of early IIn or later photospheric spectra, compared to SN 2018zd and SN 2023ixf, suggesting a more compact CSM.

\section{Insights into the early flash spectroscopy}
\label{sec:insig}

\subsection{Fluctuation of Ionization States}
\label{subsec:ion}
Within the initial 72 hr after the first light, we obtained a total of twelve spectra revealing rapid evolution of line strengths, in particular for lines associated with different ionization levels (e.g., \HeI\,vs \HeII).
During this period, the lower ionization species (such as \HeI, \CIII, and \NIII) gradually weakened and disappeared, while the strength of emission lines from higher-ionization species (like \HeII, \NIV, and \CIV) increased (see also Section \ref{subsec:72h}).

The evolution of these line fluxes or EWs is however more complex. The spectral lines such as \HeI\,$\lambda$\,6678, \HeI\,$\lambda$\, 7065, and  \CIII\,$\lambda$\,5696 disappeared at $t\sim$ 29.2 hr,  while higher ionization lines, including \CIV\,$\lambda$$\lambda$\, 5801,5811 and \OV\,$\lambda$$\lambda$\,5576,5598, emerged. However, five hours later, \HeI\,$\lambda$\,6678 reappeared, while \OV\,became more elusive. In the 38.3 hr and 38.4 hr spectra, roughly four hours afterward, \HeI\,$\lambda$\,6678 vanished once again, \OV\,reappeared, while \NIII\,significantly diminished, and \NV\,became detectable.

The weakening of \NIII\,and the emergence of \NV\,were already evident in the 29.2 hr spectrum. Considering the evolution of \OV\, and \HeI, we believe the relative intensities of \NV\,and \NIII\,observed at 29.2 hr are reliable despite the detector saturation in this wavelength region. Thus, we observed a decreasing and increasing ionization across the 29.2, 34.4, and 38.3 hr spectra. 

This trend is mirrored in Figure \ref{<FI>}, where a similar fluctuation in \CIV\,$\lambda$$\lambda$\,5801,5811 is apparent in $t\sim$ 30 h spectrum. These variations suggest that although the ionization evolution initially arose and fell within the first 72 hr, there were fluctuations when the ionization reached its maximum at around 30 hours after the first light. These fluctuations are probably related to the complicated propagation through a clumpy, and perhaps asymmetric CSM. The alteration in the ionization state indicates the complex interaction between the shockwave, the surrounding material and the high-energy radiation environment in the early stages of SN explosion.

\subsection{Comparison with Spectra Model}
\label{subsec:spmo} 
\begin{figure*}
\centering
\includegraphics[width=16cm,angle=0]{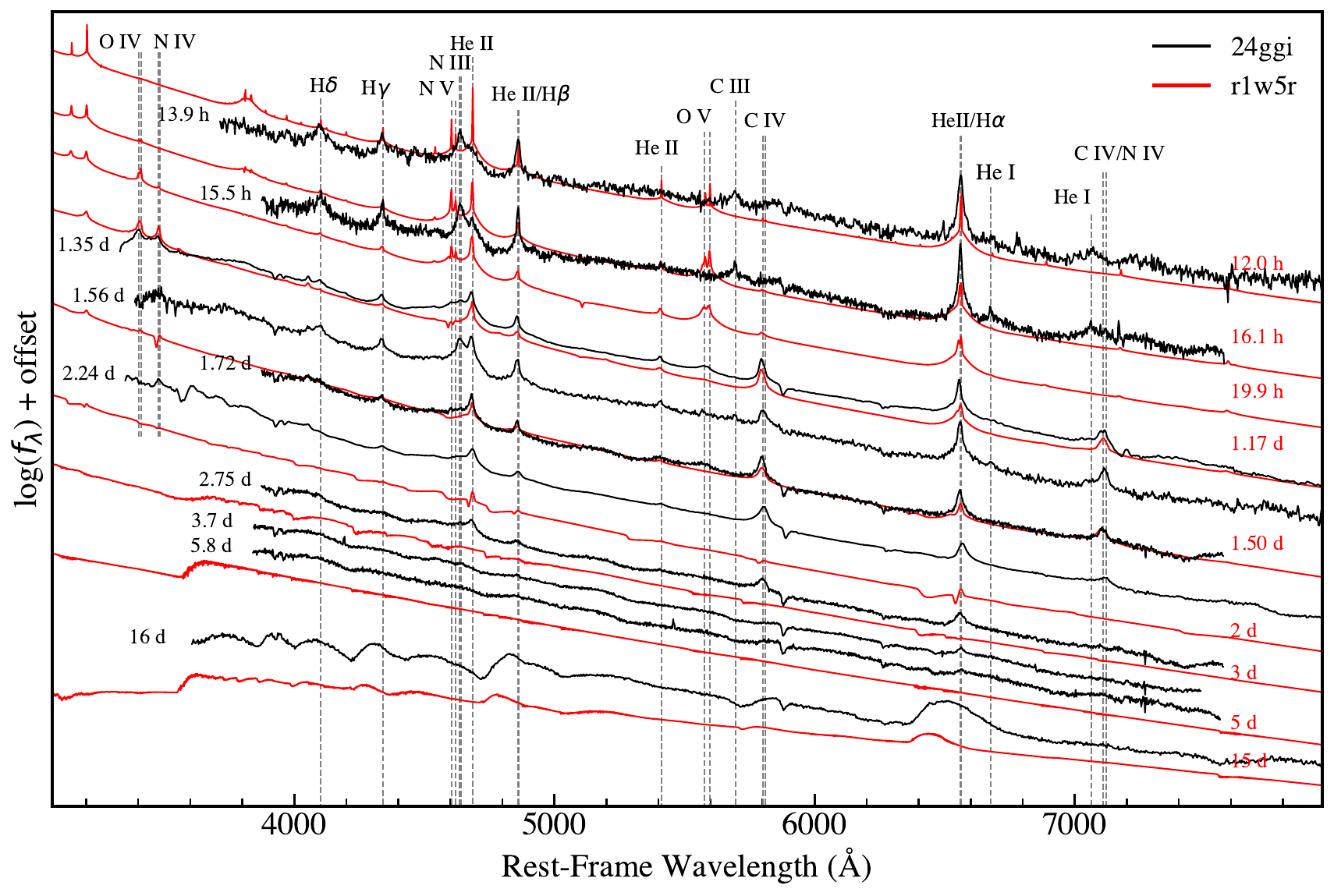}
 \caption{Comparison of early-time spectra from SN 2024ggi from 14 hr until 15 d post explosion with the r1w5r model spectra of D17. Dashed grey lines represent the rest-frame wavelengths of spectral features. In r1w5r model, the progenitor star radius is $R_{\star}=501\,{\rm R}_{\odot}$, the wind mass-loss rate is $\dot{M}=5.0 \times10^{-3}\,{\rm M}_{\odot}\,{\rm yr}^{-1}$, and  the outer radius of the dense CSM is $R_{\rm CSM} = 5 \times 10^{14} {\rm cm}$.  }
\label{<spe_D17>}
\end{figure*}

To further understand the evolution of the ionization state of SN 2024ggi, we compare the observed spectra with the spectral models presented in \cite{2017A&A...603A..51D} (referred to as D17 below), as shown in Figure \ref{<spe_D17>}. This model set successfully reproduces the observational manifestations of SNe II, specifically those featuring short-lived IIn-like spectra while taking into account the varying physical conditions of the progenitor before the explosion. The methodology adopted in D17 involves creating a comprehensive grid of models for different CSM characteristics (including radius, mass, density, density profile, velocity, and composition) based on the same progenitor and explosion parameters, thereby encompassing a wide range of the parameter space.

Although none of the models in D17 can fully reproduce the entire spectral features, models like r1w5r and r1w6 provide a good reference for us to track and study the ionization state and evolution of the CSM. In comparison, we found that the r1w5r model fits the observations well, especially at $t\sim 1.5$\,d, when it can reproduce all the observed characteristics of SN 2024ggi. Moreover, at $t\sim 1.17$ d, high-ionization lines such as \OIV\,and \NIV\,appeared in the model spectrum, which is comparable to the spectrum of SN 2024ggi at $t\sim$ 1.35 d, indicating that the ionization states of SN 2024ggi is consistent with the model during this period. However, the CSM in the r1w5r model has a higher ionization state at $t < 1$ d than that observed in SN 2024ggi. For example, the r1w5r model does not show any \HeI\,line in the early stage, and in fact No \HeI\,line appears in any of the D17 models\footnote{The reason is partly because no model was computed during the earliest epochs when the first radiation from the shock was crossing the CSM (i.e., the radiative precursor). Modeling this phase is quite challenging, and there are inconsistencies in the physics addressed in D17. One such inconsistency relates to light travel time effect, considered in radiation-hydrodynamics calculations but omitted during the post-processing phase of radiative transfer calculations for spectra.
}. In contrast, the \HeII\,line in the r1w5r model is much stronger than the \NIII\,line at the beginning of explosion.

Interestingly, the blue edge of H$\alpha$ is quite similar in the  $t\sim 16$ spectrum of SN 2024ggi and the $t = $ 15 d r1w5r model spectrum. The only difference lies in the emission strength and the extent of the profile to the red. One reason might be that in the model, the optical depth attenuation is much more significant, which suggests the CDS breaks up and becomes clumpy significantly. This could be resolved if the CDS was allowed to break-up, as would occur in multi-D radiation-hydrodynamics with the development of Rayleigh-Taylor instabilities. 
 
\cite{2024arXiv240419006J} favored the r1w6 model over the r1w5r model in their comparisons. The mass loss rates of these two models are quite similar, and their early spectra are also very similar. Both models exhibit a higher ionization in the early stages than observed in SN 2024ggi. The early \HeII\,line in the r1w6 model is stronger than that in the r1w5r model, leading us to believe that the r1w5r model performs better in this aspect. Additionally, since the spectra of \cite{2024arXiv240419006J} do not extend beyond ten days after the explosion, they did not observe the distinct P-cygni features that appeared later in SN 2024ggi. It is worth noting that the early stage spectra of the r1w6 model (up to 14 days) also do not show this characteristic, making it inferior to the r1w5r model. 

The r1w5r model is suitable for SN 2024ggi, as it can roughly reproduce the observed main IIn-like spectral features. However, the issue lies in the fact that the early ionization state of the model is too high. The CSM property of SN 2024ggi might be intermediate between r1w5r and r1w6. There may also be additional features ignored in D17, such as asymmetry, the break-up of the CDS, clumping in the CSM, and such. And the radiative transfer modeling could be improved. Therefore, the initial spectral evolution of SN 2024ggi provides a new opportunity for future developments of spectroscopic models.

\begin{figure}
\centering
\includegraphics[width=8cm,angle=0]{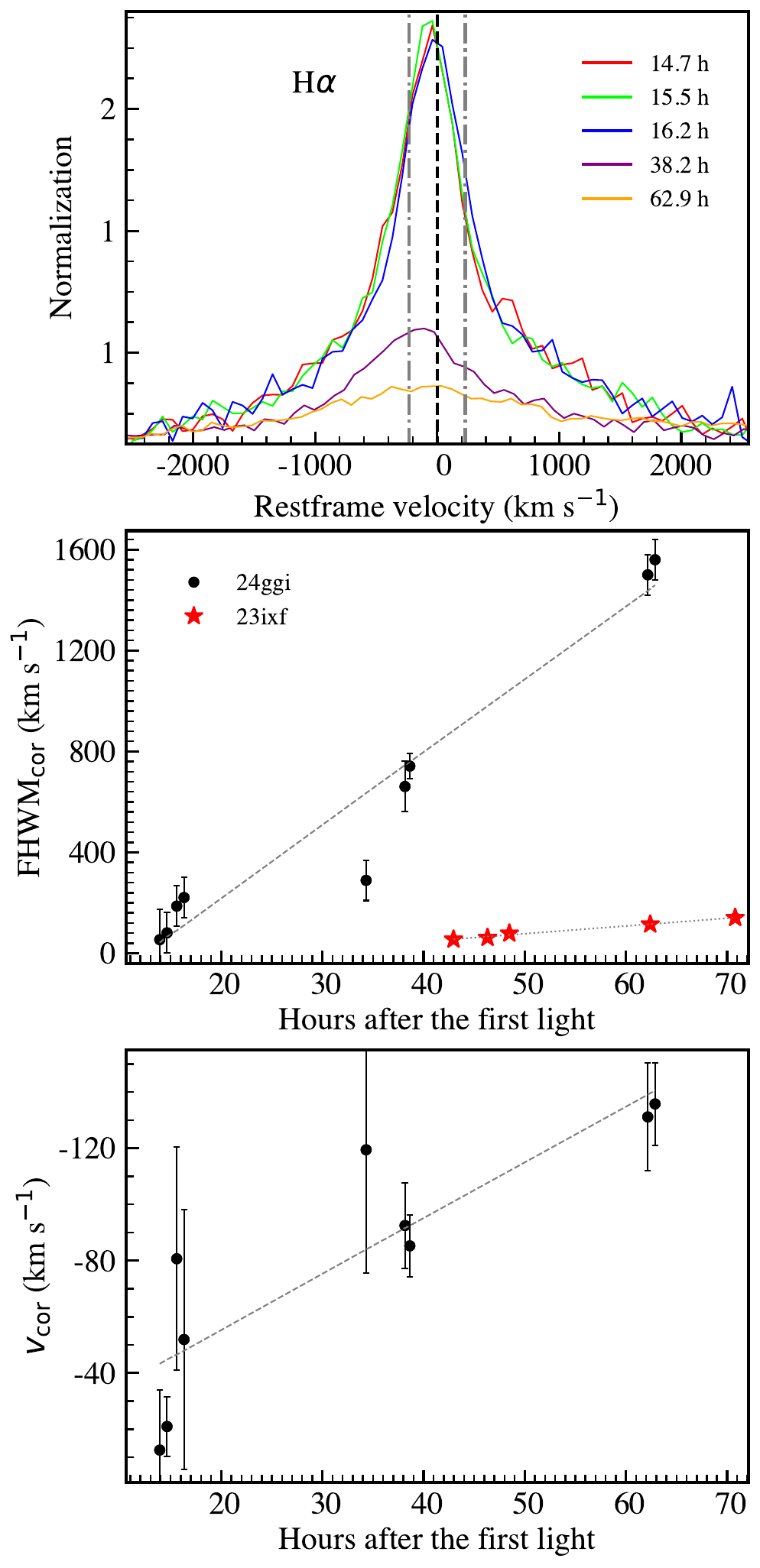}
 \caption{ Top: H$\alpha$ emission spectra of SN 2024ggi captured at $t < 3$d obtained by LJT+YFOSC(G14). The black dashed lines denote the zero-velocity mark. The instrumental FWHM  (460 \kms) of these spectra is indicated by the grey dotted-dashed line. Middle: The evolution of the corrected FWHM of the narrow  H$\alpha$ component of SN  2024ggi comparing with that of SN 2023ixf derived by the mid-resolution spectra of \cite{2023SciBu..68.2548Z} and a high-resolution spectrum (at $t \approx 62.4$ hr) of \cite{2023ApJ...956...46S}. Bottom: Velocity evolution derived from the H$\alpha$ emission of SN 2024ggi at $t<3$ d. The dashed line represents a linear fit to the velocity data. The derived velocities have been corrected for skyline \OI\,$\lambda$ 6300.3, as well as adjusted for the host galaxy's rotation at $v = -60 \pm $10 \kms, which was determined through analysis of the \NaI\,D absorption in mid-resolution spectra.  }
\label{<Haline>}
\end{figure}

\subsection{Evolution of Hydrogen emission}
\label{subsec:Ha}
In the early phase, when the SN photosphere is within the unshocked ionized CSM, the spectra are characterized by symmetric emission lines with narrow cores and extended, electron-scattering broadened wings. Figure \ref{<twocomp>} shows the double-component fitting of the early H$\alpha$ emission of SN 2024ggi. Note that the broader wings are referred as mid-width components to distinguish them from the even wider emission lines in the later P-Cygni profiles Doppler-broadened in the fast-expanding ejecta.

Based on the two-component fitting of Figure \ref{<twocomp>}, we obtain the parameters and evolution of each component. To visually compare this evolution in H$\alpha$, we selected spectra taken by YFOSC+G14 in the first three days, as presented in the top panel of Figure \ref{<Haline>}, to minimize the influence of instrumental effects. All of these spectra are corrected for the redshift and rotation of the host galaxy; see more detail in section \ref{subsec:ext}.  

Given the low spectral resolution, an instrumental correction is necessary to accurately determine the FWHM of the H$\alpha$ line via FWHM$_{\rm cor}$ = (FWHM$^2_{\rm obs}$ $-$ FWHM$^2_{\rm inst})^{1/2}$, where instrumental FWHM is estimated through the skyline \OI\,$\lambda$ 6300.3. This method has been approved effective in the study of SN 2023ixf, where the value we obtained through mid-resolution spectra \citep{2023SciBu..68.2548Z} matches well with that from the high-resolution spectra \citep{2023ApJ...956...46S}, as seen in the middle panel of Figure \ref{<Haline>}. In this panel, the corrected FWHM of the narrow component broadens from about 50 \kms\,to approximately 200\,\kms\,within hours on the first day. Initially, the FWHM$_{\rm obs}$ is close to the instrumental FWHMM$_{\rm inst}$, making corrections potentially inaccurate. We propose 50 \kms as an upper limit of FWHM before radiation broadening, and the actual FWHM is likely narrower at $t < 14$ hr.

Despite a measurement uncertainty of around 50\% in the third spectrum, there is a discernible broadening of spectral lines, indicating the onset of electron scattering effects as well as the potential radiative acceleration of the unshocked CSM (D17). At this point, using the FWHM of the narrow lines to constrain the original CSM velocity becomes unfeasible.

\cite{2024arXiv240502274P} measured an FWHM $\approx$ 55 \kms\,of H$\alpha$ at $t\sim$ 23.6 hr (the epoch is calculated by the first light adopted in this letter) with high-resolution spectroscopy. They found that this line broadened to 61 \kms\,seven hours later. Their measurements indicate that although the low-resolution results are not precise enough, they can provide certain constraints at the earliest hours.

At $t \sim $ 38.7 hr,  a mid-resolution spectrum revealed that the FWHM of H$\alpha$ had increased significantly, rendering instrumental broadening negligible. Remarkably, the narrow spectral lines of SN 2024ggi expanded to over 700 \kms\,within 48 hr. In contrast, the H$\alpha$ FWHM of SN 2023ixf during the similar timescale after the first light was only 55 \kms\,($t\sim 43$ hr, \citealp{2023SciBu..68.2548Z}), see in the mid-panel of Figure \ref{<Haline>}. Furthermore, the broadening of the H$\alpha$ narrow-line component in SN 2023ixf occurred much slower, increasing only from 55 \kms\,to 140 \kms\,within the first three days, and reaching close to 900 \kms\,when the narrow line was almost disappearing on the fifth day. Similarly to SN 2023ixf, the FWHM of the narrow H$\alpha$ of SN 2020pni increased from 250\,\kms\,to nearly 1000\,\kms\,in the first five days \citep{2022ApJ...926...20T}. The more rapid and drastic broadening observed in SN 2024ggi suggests a less extended CSM than that of SN 2023ixf and SN 2020pni.

Although we cannot use the FWHM to limit the CSM velocity of SN 2024ggi, a noticeable blueshift was observed in the H$\alpha$ emission. This shift is evident in the top panel of Figure \ref{<Haline>}, with precise measurements detailed in the middle panel. After some velocity adjustments and a refined wavelength calibration using the skyline \OI$\lambda$ 6300.3, an initial blueshift velocity of $-12 \pm 20$ \kms\,was measured at $t\sim 13.9$ hr. The resolution of the second spectrum at $t\sim 14.7$ hr is slightly higher, and the measured velocity is $-20 \pm 10$ \kms. Considering the measurement errors in galactic rotation, the lower limit of the stellar wind velocity observed at the outer layer of CSM around SN 2024ggi is $ 20 \pm 15$ \kms.  A fast expansion of narrow H$\alpha$ component was also seen in the SNe II with a short-lived IIn-like feature, e.g., SN 2020pni

We averaged the measurements daily to reduce uncertainties, yielding mean velocities of $42 \pm 29$ \kms, $99 \pm 23$ \kms, and $133 \pm 17$ \kms\, in the first three days, respectively. These measures reveal the acceleration of the unshocked CSM by the radiation from the shock. The acceleration implies that the earlier observations can more genuinely reflect the movement of CSM before the explosion, representing the progenitor's stellar wind speed in its final phase. Thus, the inferred CSM velocity of SN 2024ggi does not exceed 40 \kms\,consistent with the high-dispersion observation results (e.g., 37 \kms, \citealp{2024arXiv240518490S}). Given the lower limit estimated before, the stellar wind velocity derived by our initial observation suggests that the progenitor wind velocity of SN 2024ggi is between 20 and 40\,\kms.

\section{Discussion}
\label{sec:conc}
The initial spectral transformation and ionization processes observed in this SN provide a unique and valuable perspective into the final moments of a massive star. The rapid transition from IIn-like spectra to the photospheric phase in SN 2024ggi suggests that the fast-expanding ejecta quickly engulfed the CSM, and the CSM density decreased sharply at large distances.

Based on the preceding analysis, we have sketched the CSM of SN 2024ggi. The blue-shifted H$\alpha$ emission indicates a wind velocity of the outer CSM at $20 < v < 40$ \kms. The comparison with the D17 model suggests a mass loss rate at the inner CSM on the order of  $5 \times 10^{-3}$ to $10^{-2}$ \Msun\,yr$^{-1}$ to produce the observed IIn-like spectral features. 

The duration and strength of the narrow emission lines depend on the radius and density of the CSM, providing valuable clues for the mass-loss rate of the progenitor. The narrow emission lines with electron-scattering broadened wings of SN 2024ggi vanished approximately six days after the explosion. Based on subsequent measurements of the hydrogen P-Cygni absorption component, the maximum ejecta velocity of SN 2024ggi is 8000 \kms. Adopting this value, we can confidently infer that the distribution range of CSM does not exceed $4 \times 10^{14}$ cm. This aligns with the photosphere radius inferred from the bolometric luminosity on the sixth day, as seen in Figure \ref{<bolo>}. Considering the upper limit of the stellar wind speed of 40 \kms, it would take at least three years for the stellar wind to reach that distance, but much longer if we adopt a slowly accelerating wind.

Comparison of SN 2024ggi with other SNe II exhibiting short-lived IIn-like spectra highlights the diversity in CSM properties. SN 2024ggi stands out for the fast transition from the IIn phase to the phase when spectral lines appear broad and dominated by Doppler broadening, which is consistent with a compact surrounding CSM. For example,  SN 2024ggi has a more compact CSM than that of SN 2023ixf (e.g., with a CSM distribution region of $7 \times 10^{14}$ cm, \citealp{2023SciBu..68.2548Z,2024Natur.627..759Z}) and SN 2018zd (e.g., with a CSM distribution region of $10^{15}$ cm, \citealp{2020MNRAS.498...84Z}). 

The analysis of the H$\alpha$ emission line reveals significant broadening within 63 hours, the FWHM increases from $\sim$ 50 \kms\,to $\sim$ 1500 \kms\, which is due to the radiative acceleration of the CSM. This is why the narrow component quickly disappears.

The early spectral (and photometric, Yan et al. in prep.) evolution of SN2024ggi indicates that we caught the SN during the shock breakout. In particular, the four spectra taken between 13.9 hr and 16.2 hr showed spectra with lines of a low ionization level. Taking into account the influence of spectral resolution, it can be assumed that the spectral lines and blackbody temperature of these four spectra remained almost unchanged. Combined with the significant increase in ionization level observed in the 29.2 hr spectrum, it can be inferred that the SBO occurred during this period. The photosphere radius at this time was approximately $10^{14}$ cm. Subsequently, fluctuations in high ionization lines were observed at 29.2 hr, 34.4 hr, and 38.2 hr, which may be related to the end of SBO. The corresponding photosphere radius at this time was $1.5\,\times 10^{14}$ cm (Figure \ref{<bolo>}). In other words, the SBO of SN 2024ggi, as observed from the IIn-like spectrum, occurred within the region between  $ 1 \times 10^{14}$ to $ 1.5 \times 10^{14}$ cm.

In summary, the study of SN 2024ggi will contribute to our understanding of the late stages of stellar evolution, the pivotal role of CSM in shaping the SN observations, the intriguing diversity of SNe II, and the process of shock wave propagation in CSM.



\begin{acknowledgments}
This work is supported by the National Key R\&D Program of China with No. 2021YFA1600404, the National Natural Science Foundation of China (12173082), the science research grants from the China Manned Space Project with No. CMS-CSST-2021-A12, the Yunnan Fundamental Research Projects (grants 202201AT070069 and 202401BC070007), the Top-notch Young Talents Program of Yunnan Province, the Light of West China Program provided by the Chinese Academy of Sciences, the International Centre of Supernovae, Yunnan Key Laboratory (No. 202302AN360001). X.-F. Wang is supported by the National Natural Science Foundation of China (NSFC grants 12288102 and 1203300), and the Tencent Xplorer Prize. Y.-Z. Cai is supported by the National Natural Science Foundation of China (NSFC, Grant No. 12303054), and the Yunnan Fundamental Research Projects (Grant No. 202401AU070063). ZG is supported by the ANID FONDECYT Postdoctoral program No. 3220029. This work was funded by ANID, Millennium Science Initiative, AIM23-0001. This work has made use of the University of Hertfordshire's high-performance computing facility (\url{http://uhhpc.herts.ac.uk}). L.-Z. Wang is sponsored (in part) by the Chinese Academy of Sciences (CAS), through a grant to the CAS South America Center for Astronomy (CASSACA) in Santiago, Chile.

We acknowledge the support of the staff of the LJT, VLT, and TNG. Funding for the LJT has been provided by the CAS and the People's Government of Yunnan Province. The LJT is jointly operated and administrated by YNAO and the Center for Astronomical Mega-Science, CAS.

\end{acknowledgments}

%

\vspace{5mm}
\facilities{YAO:2.4m (LJT), TNG, VLT:Antu, Magellan:Baade.}

\software{PyRAF \citep{2012ascl.soft07011S}, NumPy \citep{2020Natur.585..357H}, Matplotlib \citep{2007CSE.....9...90H}, Astropy \citep{2013A&A...558A..33A,2018AJ....156..123A,2022ApJ...935..167A}}



\appendix
\section{Appendix}

 \subsection{Data reduction}
 \label{subsec:redu}

Figure\ref{<spe_whole>} displays the spectra of SN 2024ggi obtained by LJT, TNG and VLT UT-1, with further specifics outlined in Table \ref{Tab:Spec_log}. All of these spectra are produced in the standard way in IRAF, including precise wavelength and flux calibration, and have been corrected for telluric absorption and redshift. The wavelength was double-checked by the skylines (e.g., \OI\,$\lambda$ 5577.3, \OI\,$\lambda$ 6300.3, and \OI\,$\lambda$ 6363.8). The flux of spectra was double-checked by the SED of $ugriz-$ band photometry, as seen in Figure \ref{<bolo>}. During the spectroscopic monitoring of LJT, we got high cadence photometry quasi-simultaneously. The flux of VLT and TNG are checked with the SED interpreted by the photometry of LJT and Burst Observer and Optical Transient Exploring System (Yan et al. in prep.). Figure \ref{<bolo>} presents the bolometric light curve of SN 2024ggi based on the $ugriz$ band photometry and blackbody fitting. The temperature derived from the blackbody fitting and the photospheric radius estimated via luminosity and temperature are also plotted. We got one near-infrared spectrum from the Magellan telescope with FIRE (Folded-port InfraRed Echellette) on Apr. 24, 2024 (t$\sim$ 14 d), as seen in Figure \ref{<NIR>}, which was reduced with FIREHOST V2.0 pipeline \citep{Gagn2015FireHose}. 
Figure \ref{<twocomp>} displays two-components fitting of H$\alpha$ emission in SN 2024ggi presented in spectra obtained in the first two days.

\setcounter{table}{0} 
\renewcommand{\thetable}{\Alph{section}\arabic{table}}
\begin{table*}
\caption{Journal of Spectroscopic Observations of SN 2024ggi}
\scriptsize
\centering
\begin{tabular}{lccccc}
\hline\hline
Date (UT) & MJD & Epoch$^a$ & Range (\AA) & Spec. Res.  & Telescope+Inst.\\
Apr. 11	&	60411.608	&	13.9hr	&	3600-8800	&	410	&	LJT+YFOSC(G3)	\\
Apr. 11	&	60411.641	&	14.7hr	&	3900-7600	&	650	&	LJT+YFOSC(G14)	\\
Apr. 11	&	60411.677	&	15.5hr	&	3900-7600	&	650	&	LJT+YFOSC(G14)	\\
Apr. 11	&	60411.706	&	16.2hr	&	3900-7600	&	650	&	LJT+YFOSC(G14)	\\
Apr. 11	&	60412.248	&	29.2hr	&	3300-9200	&	370	&	VLT UT1 + FORS2(300V)	\\
Apr. 11	&	60412.463	&	34.4hr	&	3400-8100	&	410	&	TNG+LRB	\\
Apr. 12	&	60412.623	&	38.2hr	&	3900-8870	&	650	&	LJT+YFOSC(G14)	\\
Apr. 12	&	60412.641	&	38.7hr	&	3500-9500	&	3500	&	LJT+YFOSC(E9G10)	\\
Apr. 12	&	60413.141	&	50.7hr	&	3300-9200	&	370	&	VLT UT1 + FORS2(300V)	\\
Apr. 12	&	60413.151	&	50.9hr	&	3600-5100	&	1200	&	VLT UT1 + FORS2(1200b)	\\
Apr. 13	&	60413.620	&	62.2hr	&	3500-9500	&	3500	&	LJT+YFOSC(E9G10)	\\
Apr. 13	&	60413.652	&	62.9hr	&	3900-8870	&	650	&	LJT+YFOSC(G14)	\\
Apr. 14	&	60414.631	&	3.60d	&	3400-9150	&	650	&	LJT+YFOSC(G14)	\\
Apr. 14	&	60414.638	&	3.61d	&	3850-9160	&	410	&	LJT+YFOSC(G3)	\\
Apr. 15	&	60415.634	&	4.60d	&	3400-9150	&	650	&	LJT+YFOSC(G14)	\\
Apr. 15	&	60416.078	&	5.05d	&	3400-9300	&	370	&	VLT UT1 + FORS2(300V)	\\
Apr. 16	&	60416.676	&	5.65d	&	3850-9160	&	650	&	LJT+YFOSC(G14)	\\
Apr. 16	&	60416.989	&	5.96d	&	3300-9200	&	370	&	VLT UT1 + FORS2(300V)	\\
Apr. 16	&	60417.002	&	5.97d	&	3600-5100	&	1200	&	VLT UT1 + FORS2(1200b)	\\
Apr. 18	&	60418.627	&	7.60d	&	3900-7600	&	650	&	LJT+YFOSC(G14)	\\
Apr. 22	&	60422.598	&	11.57d	&	3900-8870	&	410	&	LJT+YFOSC(G3)	\\
Apr. 24	&	60424.980	&	13.95d	&	8400-24000	&	6000	&	Magellan+FIRE	\\
Apr. 25	&	60425.626	&	14.60d	&	3500-9500	&	3500	&	LJT+YFOSC(E9G10)	\\
Apr. 26	&	60426.554	&	15.52d	&	3600-8800	&	410	&	LJT+YFOSC(G3)	\\
Apr. 29	&	60429.589	&	18.56d	&	3900-7600	&	650	&	LJT+YFOSC(G14)	\\
May 03	&	60433.557	&	22.53d	&	3600-8800	&	410	&	LJT+YFOSC(G3)	\\
May 08	&	60438.566	&	27.54d	&	3600-8800	&	410	&	LJT+YFOSC(G3)	\\
May 12	&	60442.556	&	31.53d	&	3600-8800	&	410	&	LJT+YFOSC(G3)	\\
\hline
\hline
\hline
\end{tabular}

$^a${The epoch is relative to the first light date, MJD = 60411.03.}\\

\label{Tab:Spec_log}
\end{table*}

\subsection{Na I D absorption and extinctions }
\label{subsec:ext}

We observed three sets of absorption lines in the mid-resolution spectrum of SN 2024ggi. Based on their wavelength relationships, we can infer that these three sets of lines are Na~{\sc i}\,D absorption lines from different redshifts, indicating dust extinction in the line of sight direction of SN 2024ggi, as seen in Figure\ref{<Na>}. The first set at  z$\sim$0.000036 is the absorption from the Milky Way. Based on the redshift, z$\sim$0.00039, the second set of absorption lines may originate from a molecular cloud within the Milky Way at a recession velocity of approximately 120 \kms. Given the redshift of NGC 3621 (z = 0.002435 $\pm$ 0.000007, from NASA/IPAC Extragalactic Database), the third set of \NaI\,D line at z$\sim$0.002235 should be from the host galaxy with a rotation velocity at $-60 \pm 10$ \kms, where the velocity error is derived by the estimation of the skyline \OI\,$\lambda$5577.3.
 
The EW of interstellar Na~{\sc i}  D1 and D2 lines can be used to estimate the dust extinctions, e.g., the empirical relations in \cite{2012MNRAS.426.1465P}. We derived the E(B-V) of SN 2024ggi depending on the three sets of  Na~{\sc i}D absorption, as listed in Table \ref{<Na>}. The estimations from the same set of D1 and D2 are averaged. Based on this, we derived the Galactic extinction is $E(B-V)_{\rm MW} = 0.048\pm 0.009$\, mag, which is close to the result of \citet{2011ApJ...737..103S} (i.e., $E(B-V)_{\rm MW} = 0.071 $ mag). The extinction of the host galaxy and the intermediate cloud are $E(B-V)_{\rm Host} = 0.063\pm 0.005$\, mag and $E(B-V)_{\rm IMC} = 0.050\pm 0.009$\, mag, respectively. Adopted the Galactic extinction of \citet{2011ApJ...737..103S} and our host and intermediate cloud estimates, the total extinction of SN 2024ggi is $E(B-V) = 0.18\pm 0.01$ mag, which is adopted in the related calculation of this paper. This measurement is consistent with the extinction results obtained by \cite{2024arXiv240502274P}, i.e., $E(B-V) = 0.16\pm 0.02$ mag,  and \cite{2024arXiv240518490S}, i.e., $E(B-V) = 0.15\pm 0.02$ mag, through high-dispersion spectroscopic observations of the  Na~{\sc i}\,D lines.

\renewcommand{\thetable}{\Alph{section}\arabic{table}}
\begin{table*}
\caption{Parameters of Na ID lines and the Extinctions of SN 2024ggi }
\scriptsize
\centering
\begin{tabular}{lcccccc}
\hline\hline
Line$^a$ & Obs. $\lambda$ & $\Delta\lambda$ & EW &$\Delta$EW& E(B-V) & $\Delta$E(B-V)\\
 &(\AA)&(\AA)&(\AA)&(\AA)&(mag)&(mag)\\
D2$_{\rm MW}$	&	5889.91	&	0.24	&	0.30	&	0.03	&	0.053	&	0.006	\\
D2$_{\rm IMC}$	&	5892.30	&	0.21	&	0.33	&	0.01	&	0.064	&	0.000	\\
D1$_{\rm MW}$	&	5895.98	&	0.62	&	0.16	&	0.02	&	0.042	&	0.003	\\
D1$_{\rm IMC}$	&	5898.26	&	0.16	&	0.22	&	0.03	&	0.062	&	0.006	\\
D2$_{\rm Host}$	&	5903.08	&	0.28	&	0.25	&	0.01	&	0.043	&	0.001	\\
D1$_{\rm Host}$	&	5909.10	&	0.52	&	0.21	&	0.01	&	0.056	&	0.002	\\
\hline
\hline
\hline
\end{tabular}

$^a${MW represents the absorption from the Milky Way, IMC represents absorption from intermediate clouds, and Host represents absorption from the Host galaxy. }\\

\label{Tab:Na}
\end{table*}

\setcounter{figure}{0} 
\renewcommand{\thefigure}{\Alph{section}\arabic{figure}}

\begin{figure*}
\centering
\includegraphics[width=12cm,angle=0]{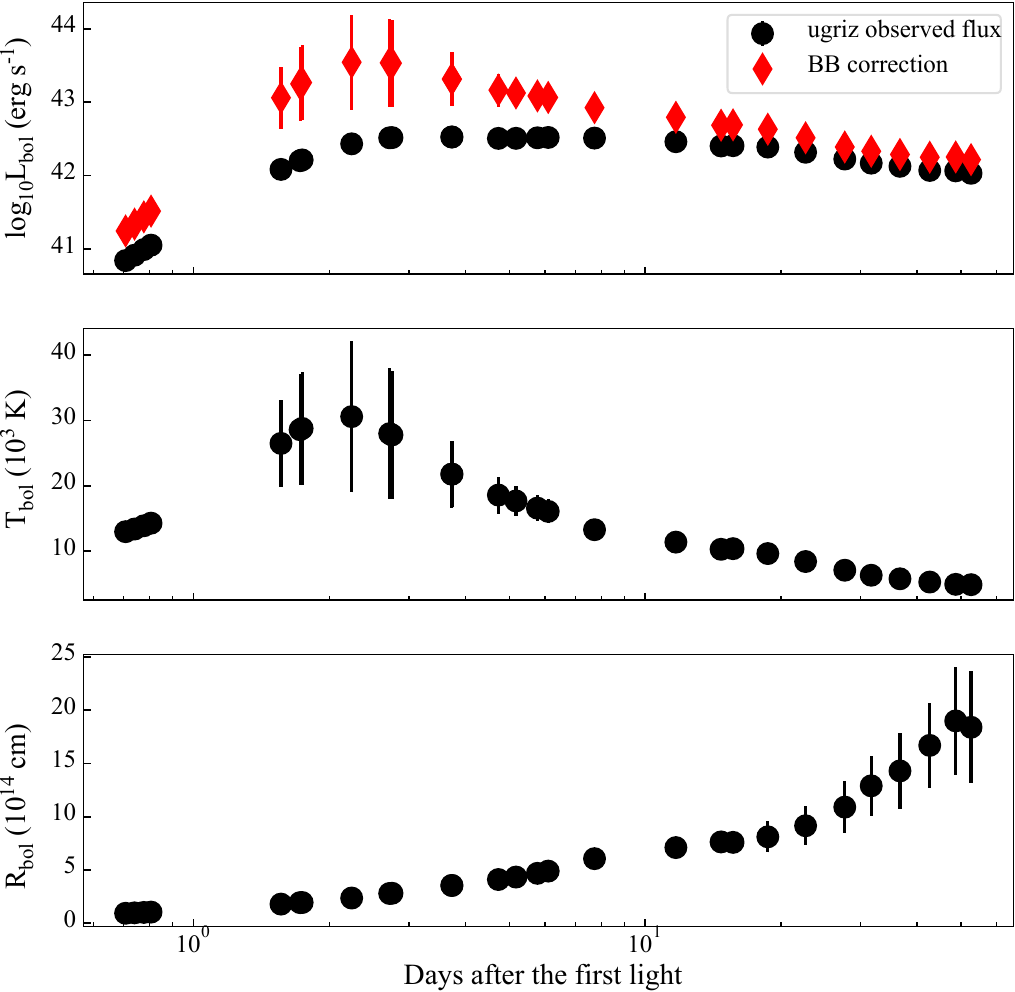}
\caption{Bolometric light curve, black body temperature, and photospheric radius calculations for SN 2024ggi based on photometric data of LJT at the same time of spectral observation. Top panel: Bolometric light curves determined through $ugriz-$ band photometry and fitted to a black body curve. Middle panel: Temperature determined from the black-body fit of the photometry data. Bottom panel: Photospheric radius determined using luminosity and temperature measurements.}
\label{<bolo>}
\end{figure*}

\begin{figure*}
\centering
\includegraphics[width=16cm,angle=0]{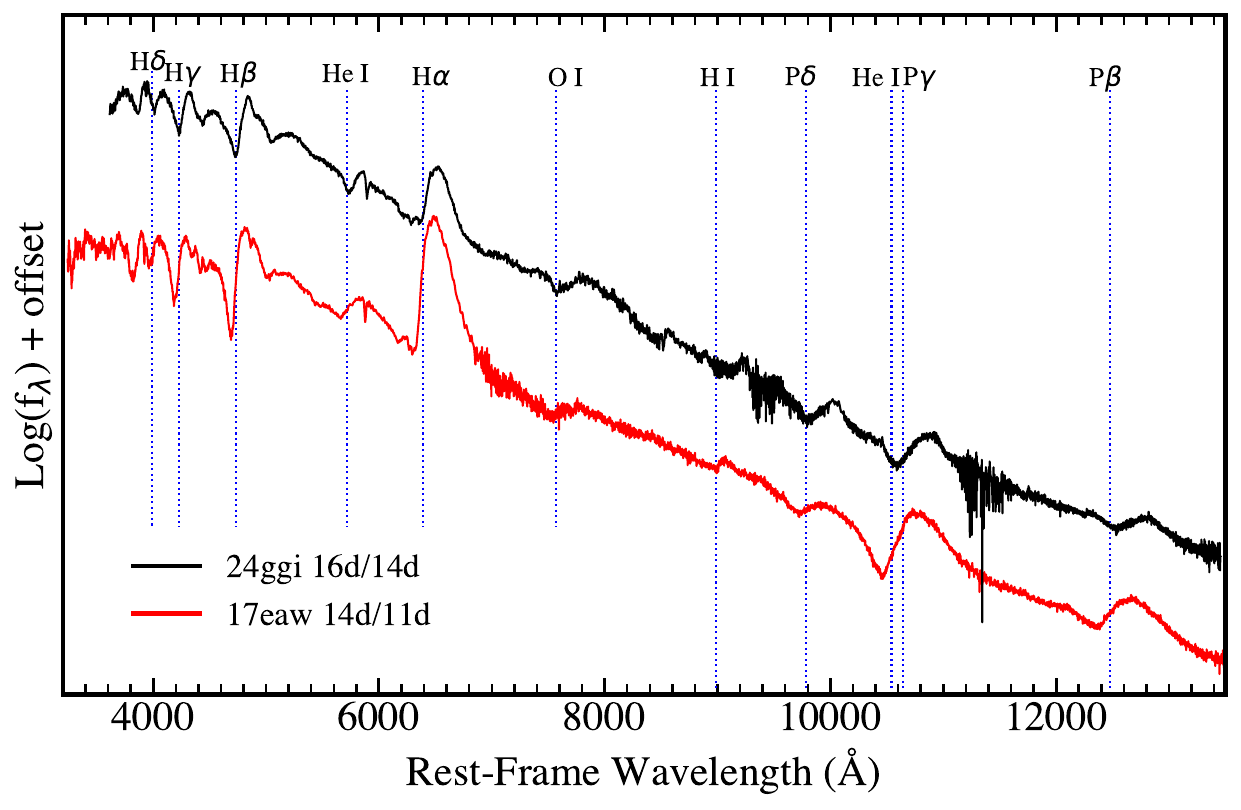}
\caption{Optical-NIR spectrum of SN 2024ggi compared with SN 2017eaw \citep{2019ApJ...876...19S}. The epoch of optical and NIR observation is marked after the name of each SN. The blue dotted line represents a blueshift corresponding to a velocity of -8000 \kms.  }
\label{<NIR>}
\end{figure*}

\begin{figure*}
\centering
\includegraphics[width=16cm,angle=0]{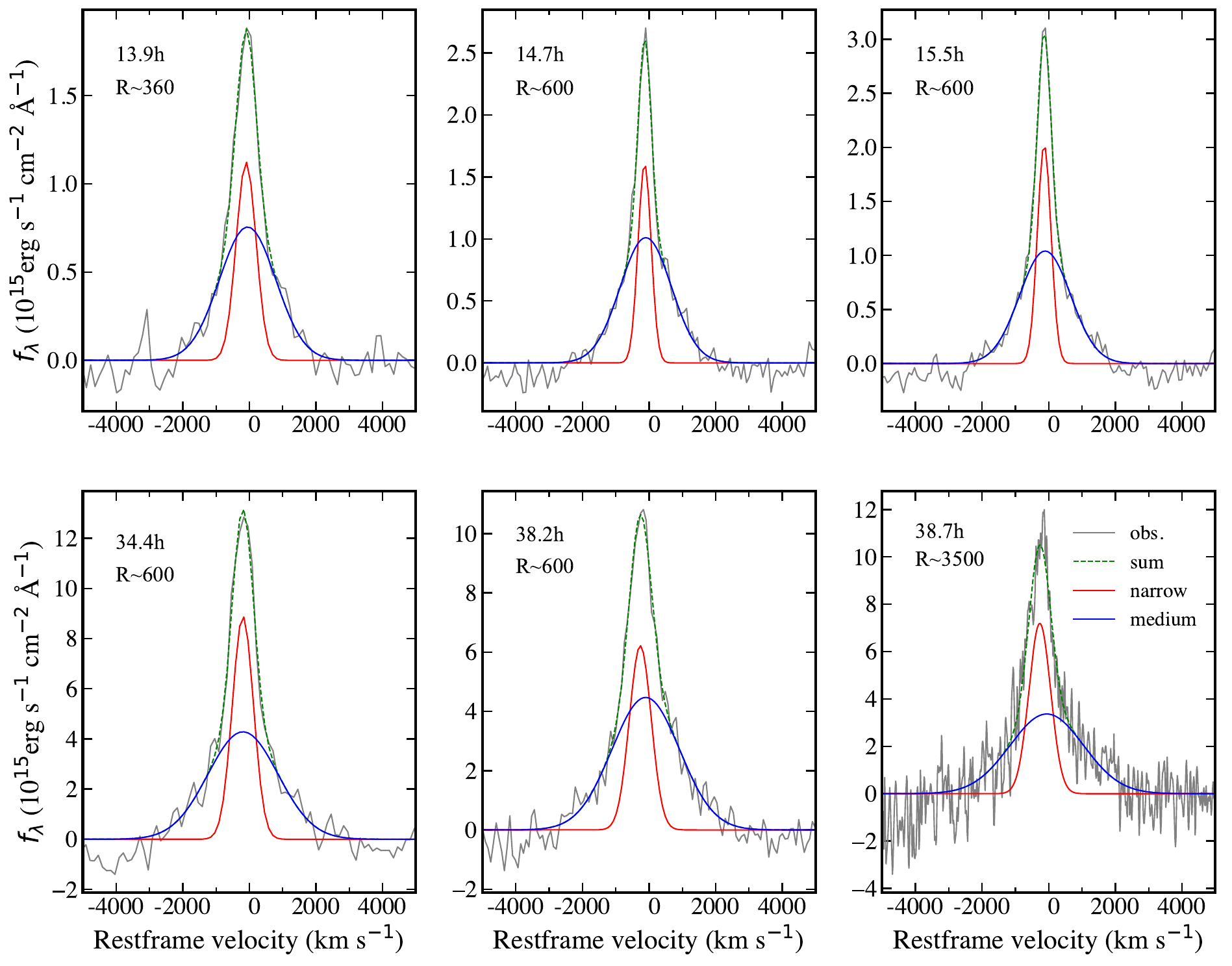}
\caption{Gaussian and Lorentzian fit the narrow and medium components of the H$\alpha$ line in the spectra of SN 2024ggi at six selected phases. The spectral resolution of each spectrum is marked.}
\label{<twocomp>}
\end{figure*}

\begin{figure*}
\centering
\includegraphics[width=14cm,angle=0]{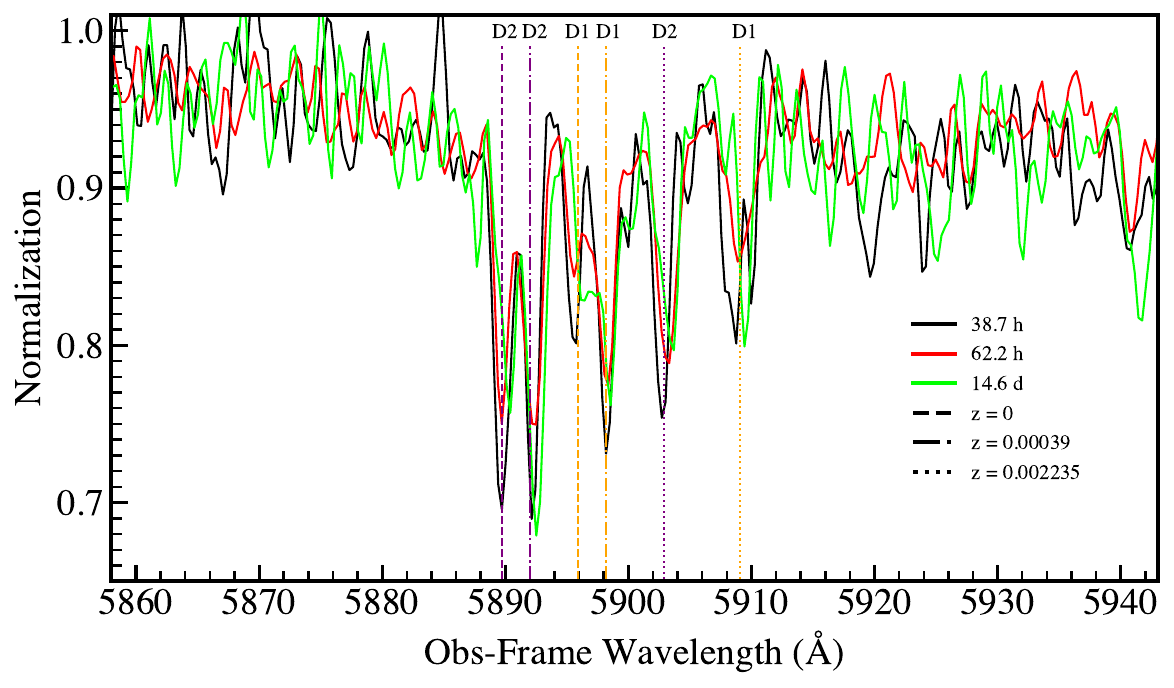}
\caption{Three set of \NaI\,D absorption in the middle resolution spectra of SN 2024ggi. The dashed, dash-dotted, and dotted lines represent the \NaI\,D at different redshifts as marked in the legend.  }
\label{<Na>}
\end{figure*}

\bibliography{ms}{}
\bibliographystyle{aasjournal}




\end{document}